\documentclass[a4paper,11pt]{article}
\usepackage{pos}

\newcommand{\arxiv}[2]{[arXiv:\,\href{http://arxiv.org/abs/#1}
{\texttt{#1}}[\texttt{#2}]]}

\title{Photon production rate from Transverse-Longitudinal ($T-L$) mesonic correlator on the lattice}

\author*[a]{D. Bala}
\author[a,b]{S. Ali}
\author[c]{A. Francis}
\author[d]{G. Jackson}
\author[a]{O. Kaczmarek}
\author[a]{T. Ueding}

\affiliation[a]{Fakult\"at f\"ur Physik, Universit\"at Bielefeld, D-33615 Bielefeld, Germany}
\affiliation[b]{Government College University Lahore, Department of Physics, Lahore
54000, Pakistan}

\affiliation[c]{Institute of Physics, National Yang-Ming Chiao Tung University, 30010 Hsinchu, Taiwan}

\affiliation[d]{Institute for Nuclear Theory, Box 351550, University of Washington, Seattle, WA 98195-1550, United States}

\abstract{
Thermal photons from the QGP provide important information about the interaction among
plasma constituents.
The photon production rate from a thermally equilibrated system is proportional to the transverse spectral function $\rho_T(\omega=|\vec k|, \vec k)$.
One can also calculate the photon production rate from the difference 
between $\rho_T(\omega,\vec k)$ (transverse) and $\rho_L(\omega,\vec k)$ (longitudinal)
projections, 
as $\rho_L$ vanishes on the photon point.
Because 
the UV part of $\rho_T-\rho_L$ is suppressed, the corresponding Euclidean correlator receives most of its contribution from the IR part. We calculate the $T\!-\!L$ correlator on $N_f=2+1$ flavour HISQ configurations with $m_l=m_s/5$ at temperature of about $1.15\,T_{pc}$ (220 MeV). 
We have used two ans\"{a}tze for the spectral function: 
1) A polynomial connected to the UV region consistent with OPE expansion and 2) a hydro-inspired spectral function. 
We have also applied 
the Backus-Gilbert method to estimate the spectral function. 
All these different approaches are combined to estimate the photon production rate.
}

\FullConference{%
  The 39th International Symposium on Lattice Field Theory (Lattice2022),\\
  8-13 August, 2022 \\
  Bonn, Germany 
}

\begin{document}
\maketitle
\section{Introduction}
Lattice QCD determinations of the 
equation of state (EoS) 
confirm the transition from hadronic matter to the quark-gluon plasma (QGP) phase under increasing temperature \cite{AB, BFHKKS}. 
Experimentally this QGP phase has been studied in high-energy heavy ion collisions at the RHIC and LHC facilities. Photons and
dileptons 
produced in the plasma are 
a vital probe 
of the QGP 
because the mean free paths of the photon and the dilepton
(i.e. virtual photon) 
are much larger than the typical size of the plasma \cite{AA}. Therefore photons and dileptons can transfer information about the plasma from the point at which it was created
and throughout its spacetime evolution. 

The photon and dilepton production rate from a 
thermalized 
QGP, at temperature $T$, to leading-order in the electromagnetic coupling $\alpha_{\rm em}$ are given by \cite{MT},
\begin{equation}
        \frac{d\Gamma_{\gamma}}{d^3{\vec k}} \; = \;
        \frac{\alpha_{\rm em} n_{B}(k)}{\pi^2 k} \;
        \Big( {\textstyle\sum_{i=1}^{N_f} Q_i^2} \Big) \;
        \rho_{T}(k,\vec k) \, , 
\end{equation}
\begin{equation}
        \frac{d\Gamma_{l^+ l^-}}{d{\omega}d^3{\vec k}} 
        \; \simeq \; 
        \frac{\alpha_{\rm em}^2 n_{B}(\omega)}{3\pi^2(\omega^2 -k^2)} \;
        \Big( {\textstyle\sum_{i=1}^{N_f} Q_i^2} \Big) \;
        \big(\,2\rho_{T}(\omega,\vec k)+\rho_{L}(\omega,\vec k)\,\big) \, ,
\end{equation}
where $n_B$ is the Bose distribution and $Q_i$ is the charge of a quark of flavour $i$
in units of the electron charge. 
All QCD information is encoded in 
$\rho_{T}(\omega ,\vec k)$ and $\rho_{L}(\omega ,\vec k)\,$ which denote the transverse and longitudinal spectral functions at frequencies $\omega$ and momenta $\vec k\,$. 

Perturbation theory is an important tool to calculate the spectral function in the weak coupling limit. One can use naive (NLO) perturbation theory away from the light cone to calculate the spectral function directly~\cite{MLAINE}. However, near the light cone, resummation 
(LPM) 
of specific Feynman diagrams is essential to obtain correct results~\cite{AMY}. 
In general, one must interpolate between the two regimes~\cite{INTERP}\footnote{
These calculations have now been extended to finite baryon chemical potential~\cite{GJ}.}. 

On the lattice, however, we need to calculate spectral functions from a lattice correlator by using the following relation,
\begin{equation}
G_{E}(\tau, \vec k) \; = \;
\int_{0}^{\infty} \frac{d\omega}{\pi} \; \rho(\omega,\vec k) \; \frac{\cosh[\omega(1/2T-\tau)]}{\sinh(\omega/2T)} \, .
\label{unstabel}
\end{equation}
Extraction of the spectral function from noisy lattice data, using the above equation, is a well-known numerically ill-posed problem. Physically motivated assumptions are required to extract the spectral function from the lattice correlator.

Following Ref.~\cite{CHMST}, we estimate the photon production rate from the $G_H=2(G_{T}-G_{L})$ correlator rather than $G_{T}$. This is because the spectral function $\rho_H=2(\rho_T-\rho_L)$ is identical to $\rho_T$ at $\omega=k$ (photon point), and the UV part of $\rho_H$ is much suppressed compared to the IR part of the spectral function. Therefore the resultant lattice correlator $G_H$ is mostly dominated by
the important IR part of the spectral function, in contrast to the $G_{T}$ correlator, which is mostly dominated by the UV part of $\rho_T$. 
Previously, 
the photon production rate has also been estimated from the correlator, which has a large UV contribution~\cite{DFKKLS, GKLM}\footnote{Recently, the photon production rate has also been estimated from the $G_{T}$ correlator \cite{CHKMT}. In Ref.\cite{TCHKMR} photon production rate has also been estimated from imaginary
momentum correlators }. It is known that the resulting $T-L$ spectral function also satisfies the following sum rule in the chiral limit~\cite{BFHMS},
\begin{equation}
    \int_{0}^{\infty} d\omega\, \omega \; \rho_{H}(\omega,\vec k) \; = \; 0 \, .
\label{sumrule}
\end{equation}

Our aim is then to compute the $G_{H}$ correlator on the lattice and make ansatz for the spectral function to fit the lattice data such that it satisfies the above sum rule.
This proceedings contribution is organized as follows. In the next section, we will present lattice details. In Sec.~\ref{spectral_reconstruction}, reconstruction of the spectral function is performed using two models and the Backus-Gilbert (BG) 
method. Sec.~\ref{FR} contains the final results of the effective diffusion coefficient (directly related to the photon production rate) extracted from the spectral functions from Sec.~\ref{spectral_reconstruction}. 

\section{Lattice Details}
\label{LD}
To calculate the correlation functions, we use $N_f=2+1$ flavor QCD  gaugefield configurations generated with the HISQ action by the HotQCD collaboration. These configurations correspond to an unphysical pion mass $m_{\pi}=315$~MeV with $m_{l}=m_{s}/5$. We used the lattice size $96^3\times 32$, corresponding to a temperature of $220$~MeV. On these configurations, we calculate the correlation function using 
clover-improved Wilson fermions 
with the parameters 
$\kappa=0.13515$ 
and tadpole improved tree-level 
$c_{sw}=1.34108$. We have used around 1700 configurations for the computation of these correlators. These parameters for Wilson clover fermion are chosen to reproduce the above staggered pion mass. 
The possible spatial momentum 
is fixed by the aspect ratio of the lattice, which is, in this case, $k_{n}/T=2\pi n N_t/N_s=2\pi n/3$ where $n$ is an integer.

\begin{figure}[hbt!]
\centering
\includegraphics[width=0.49\textwidth]{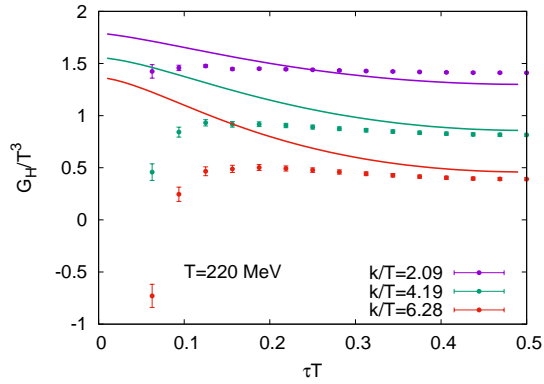}
\caption{The Euclidean hadronic correlator $G_H$ as measured on the lattice, at the points
$\tau T = \{ \frac{2}{32}, \frac{3}{32} , \ldots, \frac{16}{32}\}$, for various momenta (namely $k_n$ for $n=1,2,3$). The continuous lines are the corresponding estimates from resummed perturbation theory (NLO+LPM) as described in Ref.~\cite{JL}.
}
\label{Latt-pert-corr}
\end{figure}
  
\section{Lattice correlator and spectral reconstruction}
\label{spectral_reconstruction}
The $G_{H}$ correlator calculated on the lattice is multiplicatively renormalizable. However, the ratio $G_{H}/2\chi_{q}T$ does not need renormalization. Here $\chi_q$ is the quark number susceptibility. On the lattice, we can calculate the correlation function $G_{\mu\nu}$ of electromagnetic current $J_{\mu}$, from which one can easily calculate this ratio by using the following formula,
\begin{equation}
     \frac{G_{H}(\tau,\vec k=(0,0,k_z))}{2 \chi_q T}= \frac{G_{xx}(\tau,\vec k)+G_{yy}(\tau,\vec k)-2(G_{zz}(\tau, \vec k)-G_{00}(\tau, \vec k))}{2 \, G_{00}(\tau, \vec k)} .
\end{equation}

The renormalized lattice correlator is obtained by multiplying this ratio by $2\chi_q$. 
We have estimated $\chi_q \approx 0.872\,T^2$ from the  order $g^6 \ln(g)$ perturbative calculation~\cite{AV}. Therefore we are essentially renormalizing the correlator perturbatively. The resulting perturbatively renormalized correlator is shown in Fig.~\ref{Latt-pert-corr}. In this figure, we have also shown the NLO+LPM perturbative estimate of the corresponding correlator~\cite{JL}. The difference between the perturbative and lattice estimates of the correlator signifies non-perturbative effects. 

As mentioned, spectral reconstruction from lattice using Eq.~(\ref{unstabel}) is an ill-posed problem. Therefore various possible spectral functions 
can reproduce 
the lattice data and well-motivated assumptions are required 
in practice for spectral reconstruction. 
Here we will present two physically motivated 
models of the spectral function that can fit our lattice data.

\subsection{Model spectral functions}
For the first model to be considered, we have used a spectral function which has the following 
piecewise polynomial form
($\omega_0$ is a parameter of the model) 
\begin{equation}
     \rho_{H}^{\rm poly}(\omega) \; = \; 
     \rho_< \, \Theta(\omega_0 - \omega) +
     \rho_> \, \Theta(\omega - \omega_0) \, , \label{poly-ansatz}
\end{equation}
where
\begin{align}
 \rho_< &\equiv
 \frac{\beta \omega^3}{2\omega_0^3}\left(5-3\frac{\omega^2}{\omega_0^2}\right)
     -\frac{\gamma \omega^3}{2\omega_0^2}\left(1-\frac{\omega^2}{\omega_0^2}\right)
     +\frac{\delta \,\omega}{\omega_0} \left(1-\frac{\omega^2}{\omega_0^2}\right)^2 \, ,\label{lw0}\\
\rho_> &\equiv
     -\frac{\beta\, \omega_0^4}{7\,\omega^4}\left (54 \frac{\omega_0^4}{\omega^4}\,-\,94 \frac{\omega_0^2}{\omega^2}\,+\,33\right)
    +\frac{\gamma \, \omega_0^5}{140\,\omega^4}\left (-81 \frac{\omega_0^4}{\omega^4}\,+\,92 \frac{\omega_0^2}{\omega^2}\,-\,11\right)
    -\frac{16\,\delta\,\omega_0^4}{35\,\omega^4}\left(1-\frac{\omega_0^2}{\omega^2}\right)^2 \, ,
\label{gw0}
\end{align}
and $\Theta$ is the Heaviside step function.
The part of the spectral function for $\omega<\omega_{0}$ is taken from Ref.~\cite{GKLM} and has been used based on the smoothness of the spectral function across the light cone.
The second form of the expansion is motivated by the operator product expansion, according to which $\rho_{H}(\omega)\sim 1/\omega^4$ at large $\omega$ \cite{BFHMS}.

The various coefficients appearing 
in Eq.~(\ref{lw0}) and Eq.~(\ref{gw0}) have been chosen such that $\rho_H(\omega)$ is smooth at $\omega=\omega_0$ and satisfies the sum rule in Eq.~(\ref{sumrule}). The parameters $\beta$ and $\gamma$ correspond to the value of the spectral function and its derivative, respectively, 
at $\omega_0$. The parameter $\delta$ controls the slope of the spectral function at the origin. We perform the fit of our data with respect to $\beta$, $\gamma$, and $\delta$, under the constraints $\delta\geq0$, $\rho_{H}(k,\vec k)\geq0$ and $\frac{\partial G_{H}(\tau) }{\partial \tau}\leq0$. The first constraint comes from the fact that in hydrodynamics, $\delta$ is related to the diffusion coefficient and, therefore, cannot be negative. The second constraint amounts to the fact that the photon production rate cannot be negative.
\begin{figure}[hbt!]
\centering
\includegraphics[width=0.49\textwidth]{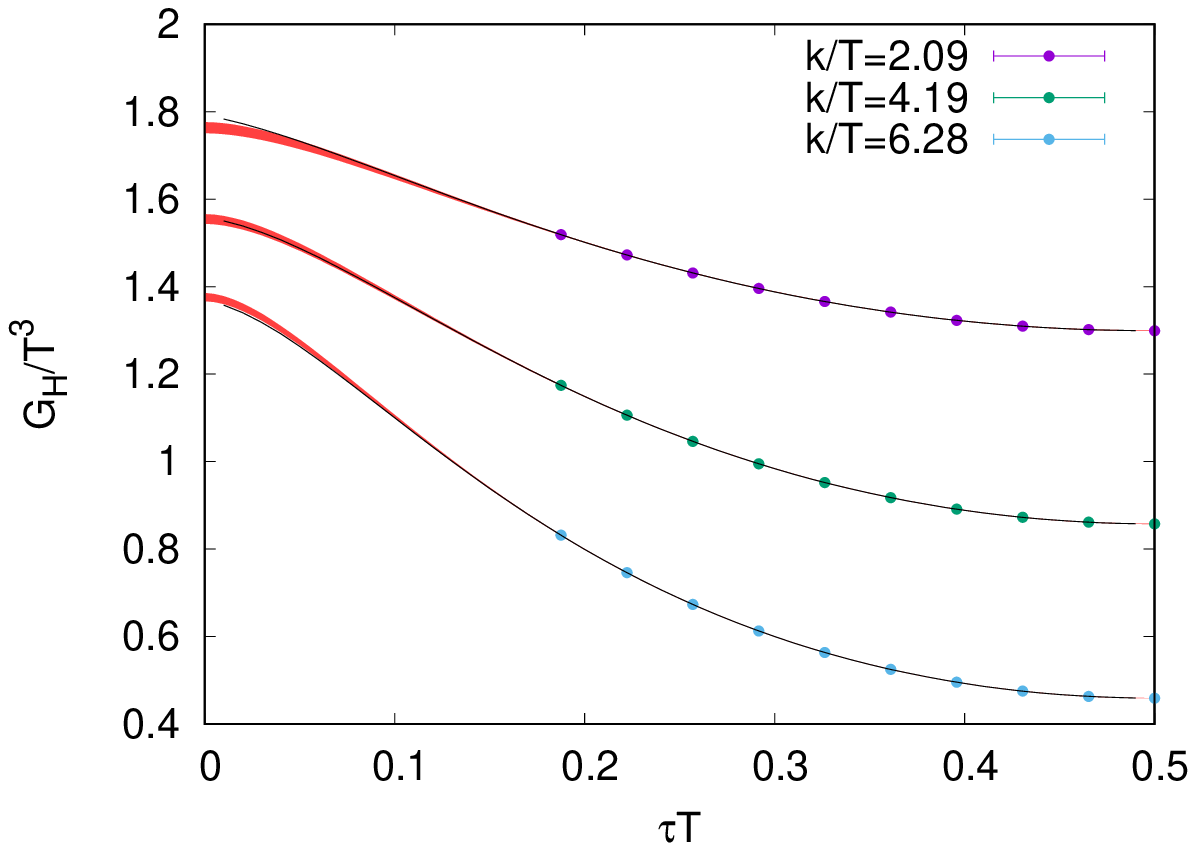}
\includegraphics[width=0.49\textwidth]{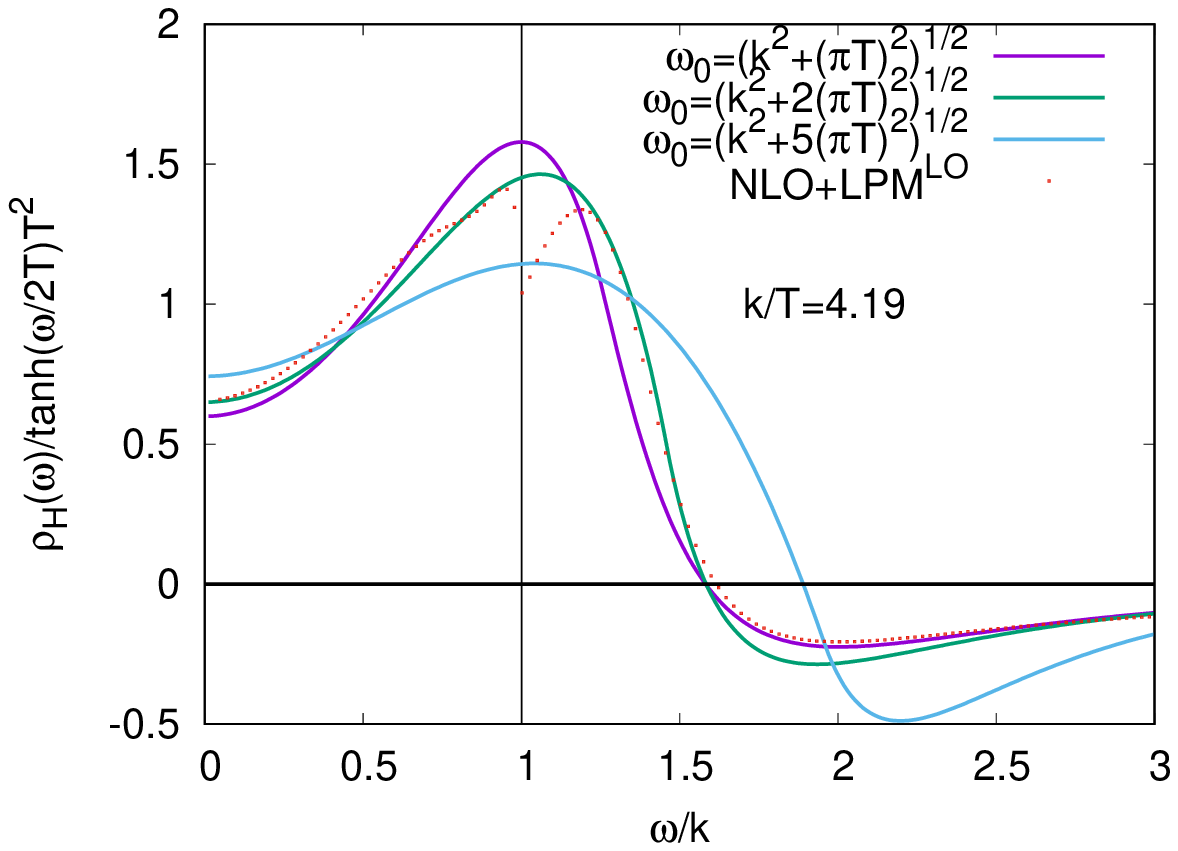}
\caption{Mock analysis of perturbative data, showing data points from 
NLO+LPM 
correlators used for the spectral reconstruction, 
the error on each point is $0.1$\% (left). 
Red bands are the correlator obtained from the fitted spectral function, 
with the width of the band stemming from uncertainty in the fit parameters. 
The black curves are the perturbative correlator at all $\tau T$ 
and thus pass through the mock data points. 
On the right, we show the reconstructed spectral function from `perturbative data,' alongside the exact spectral function as a red dotted curve.
}
\label{Mock-analysis}
\end{figure}

Before fitting the above spectral ansatz to the lattice data, we performed a mock data analysis 
wherein we reconstruct the exactly known NLO+LPM spectral function from the known NLO+LPM correlation function. 
To perform the fit with above the spectral function, we have artificially introduced an error into the perturbative correlator of the following form $\frac {\delta G_{H}}{G_{H}}=0.001$. For this reconstruction, we used 10 data points between $\tau T=0.1875$ to $\tau T=0.5$. The resulting spectral function, along with the fitted correlator, is shown in Fig.~\ref{Mock-analysis}.
In the left panel of Fig.~\ref{Mock-analysis}, we depict the perturbative data points used for the spectral reconstruction. 
The red band corresponds to the correlator obtained from the fitted spectral function, and the black curve is the perturbative correlator. We see that we could predict data points beyond the fitting range with the above-fitted ansatz.
On the right panel of Fig.~\ref{Mock-analysis}, the reconstructed spectral function for a few values of the matching point $\omega_0$ is shown. The red dotted line corresponds to the exact spectral function for the correlator. We observe that the spectral function shows some change with respect to $\omega_0$, however within a systematic uncertainty of $\omega_0$ between $\sqrt{k^2+(\pi T)^2}$ and $\sqrt{k^2+5(\pi T)^2}$ we could indeed be able to 
encompass the exact spectral function. 
\begin{figure}
\centering
\includegraphics[width=0.49\textwidth]{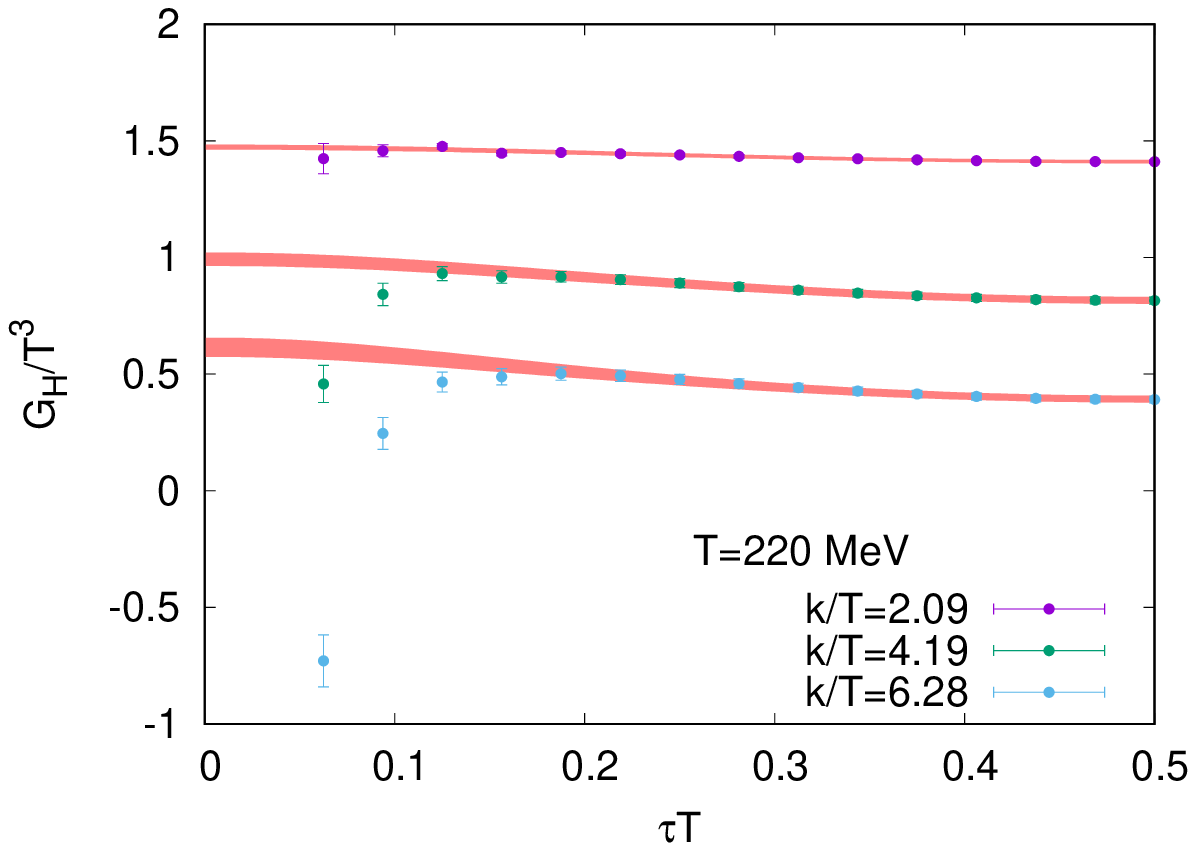}
\includegraphics[width=0.49\textwidth]{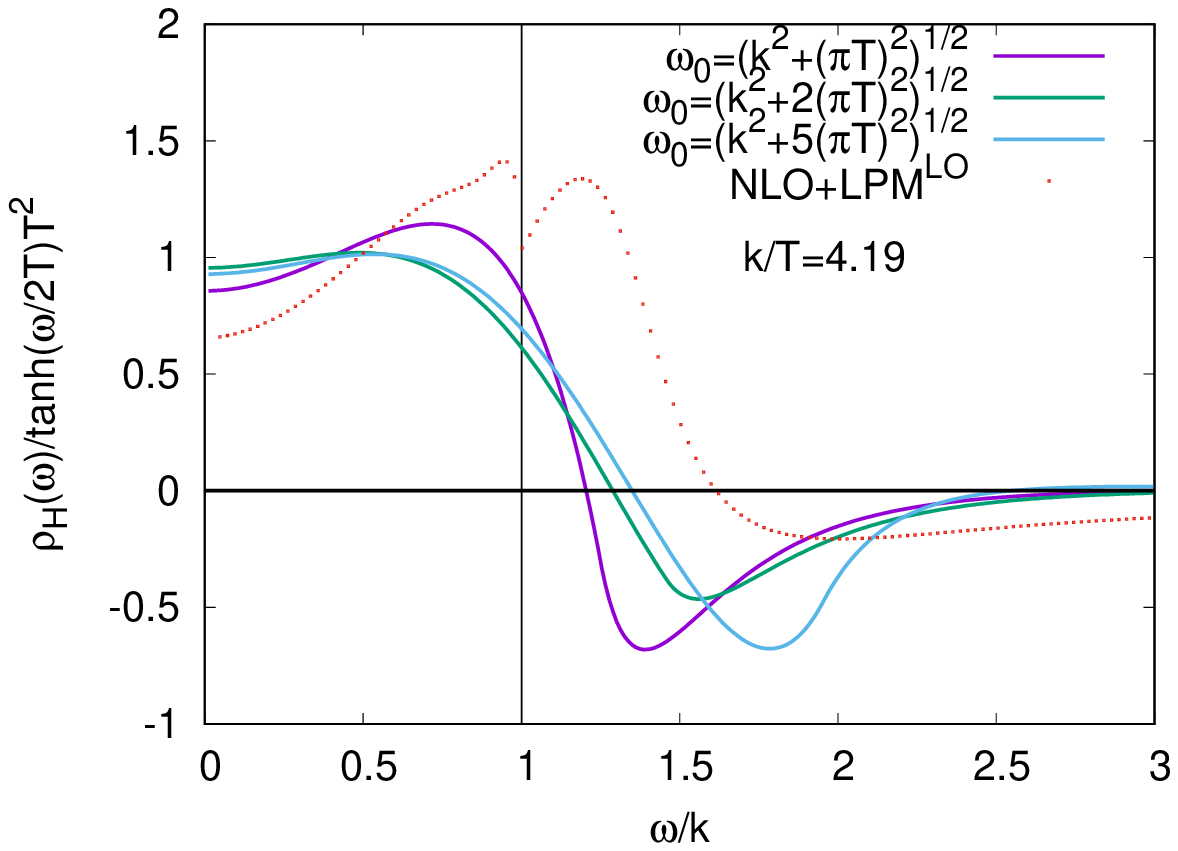}
\caption{(Left) The data points correspond to the lattice correlator calculated at various momenta. The red band is the correlator from the fitted spectral function.
(Right) The reconstructed spectral function for $k/T=4.19$ along with the
perturbative spectral function. }
\label{lattice-corr-sp}
\end{figure}

After verifying the above spectral function with the perturbative data, we now proceed to fit the lattice correlator. The fitting range we have taken is in $\tau T$ between $0.1875$ and $0.5$, as the short distance part is dominated by lattice artifacts. The spectral function for various $\omega_{0}$ is shown in the right panel of Fig.~(\ref{lattice-corr-sp}). The red shaded band in the right panel corresponds to the fitted correlator. We see the expected feature that the spectral function is IR-dominated. For the calculation of photon production rate, we take the variation with $\omega_{0}$ as a source of systematic error from this ansatz. We have also shown the corresponding NLO+LPM perturbative spectral function in the same figure. We see that at $\omega\ll k$, the perturbative spectral function is suppressed compared to our model spectral function, whereas near the light cone, the model spectral function 
is consistently below the value of the perturbative prediction. 

\begin{figure}
\centering
\includegraphics[width=0.49\textwidth]{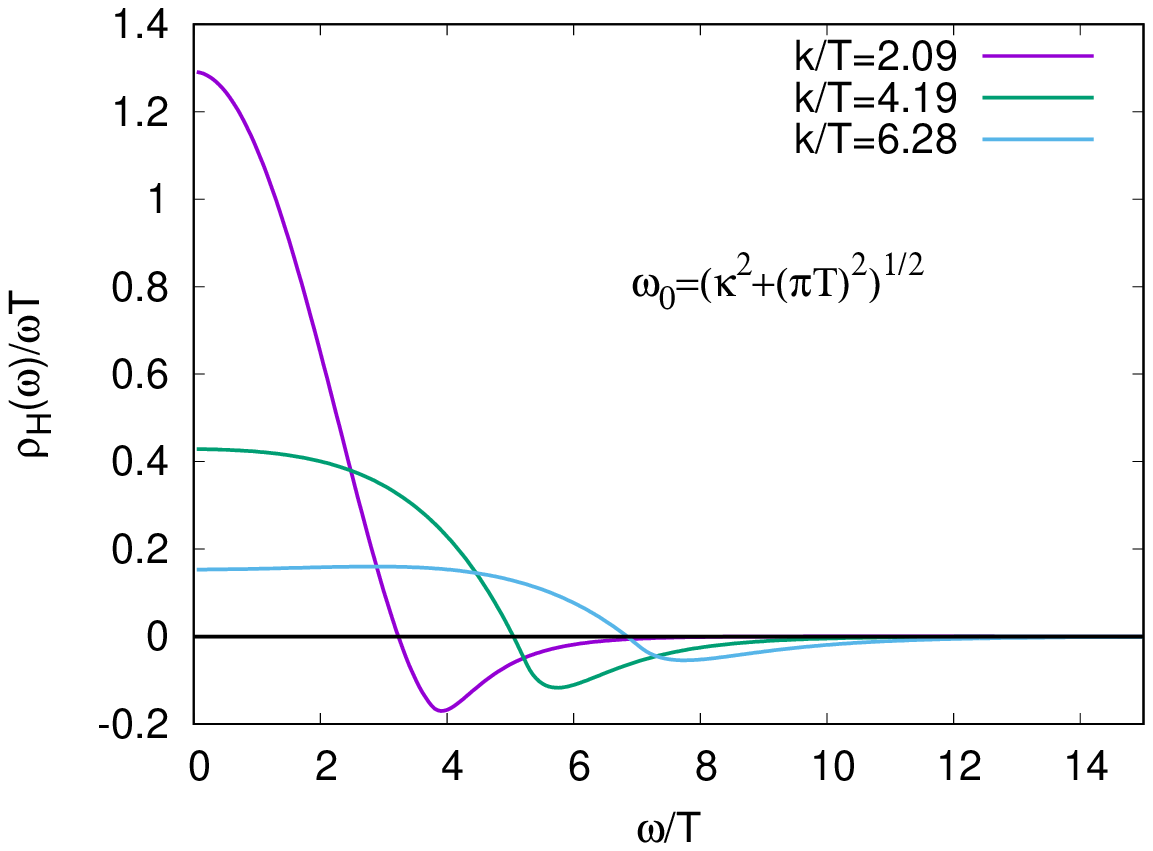}
\includegraphics[width=0.49\textwidth]{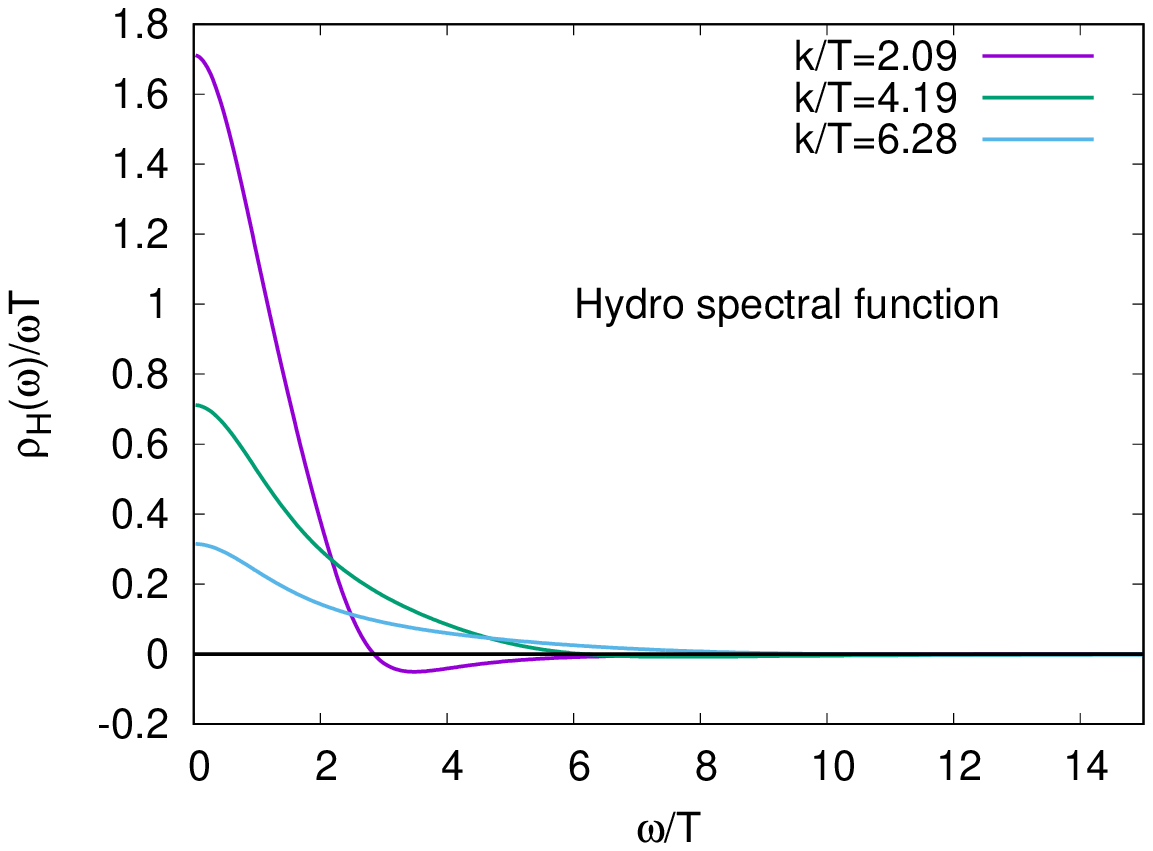}
\caption{Spectral function for various momenta, as a function of $\omega$.
In the left panel, we show the result from a polynomial ansatz (\ref{poly-ansatz}), and the right panel depicts the best fit from a hydro ansatz (\ref{hydro-ansatz}). Both functions have the correct UV behavior. 
}
\label{spectral-function-variation-M}
\end{figure}
We show the spectral function for $\omega_{0}=\sqrt{k^2+(\pi T)^2}$ for various momenta in Fig.~(\ref{spectral-function-variation-M}). We observe  that the IR  part of the spectral function is suppressed for higher momentum compared to small momentum. This behavior is expected because the spectral function becomes $\omega \,\delta(\omega)$ by rotational symmetry at zero momentum. At finite momentum, this $\delta (\omega)$ function develops a finite width, which increases with increasing momentum.

The second ansatz is motivated by relaxation-time hydrodynamics~\cite{FGR}. The spectral function in this model is 
taken as 
\begin{equation}
        \rho^{\rm hydro}_{H}(\omega,\vec k) \; = \;
        A \frac{\tanh(\omega/T) \, k^2\,(1+a^2\, k^2-2\, a\, b\, \omega^2+b^2\omega^2)}{(1+b^2 \,\omega^2)(a^2\,k^4+\omega^2-2 a\, k^2 \,b\,\omega^2+b^2\, \omega^4)} \, ,
        \label{hydro-ansatz}
\end{equation}
with parameters $A$, $a$ and $b\,$.
At large $\omega$, this spectral function goes like $1/\omega^4$ 
in accordance with the 
OPE prediction. The sum rule, on the other hand, will allow the parameter $a$ to be expressed in terms of $b$. Then we perform a two-parameter fit of this spectral function with the lattice data. The resulting spectral function for various momenta has been shown in the right panel of Fig.~\ref{spectral-function-variation-M}.
\begin{figure}
\centering
\includegraphics[width=0.49\textwidth]{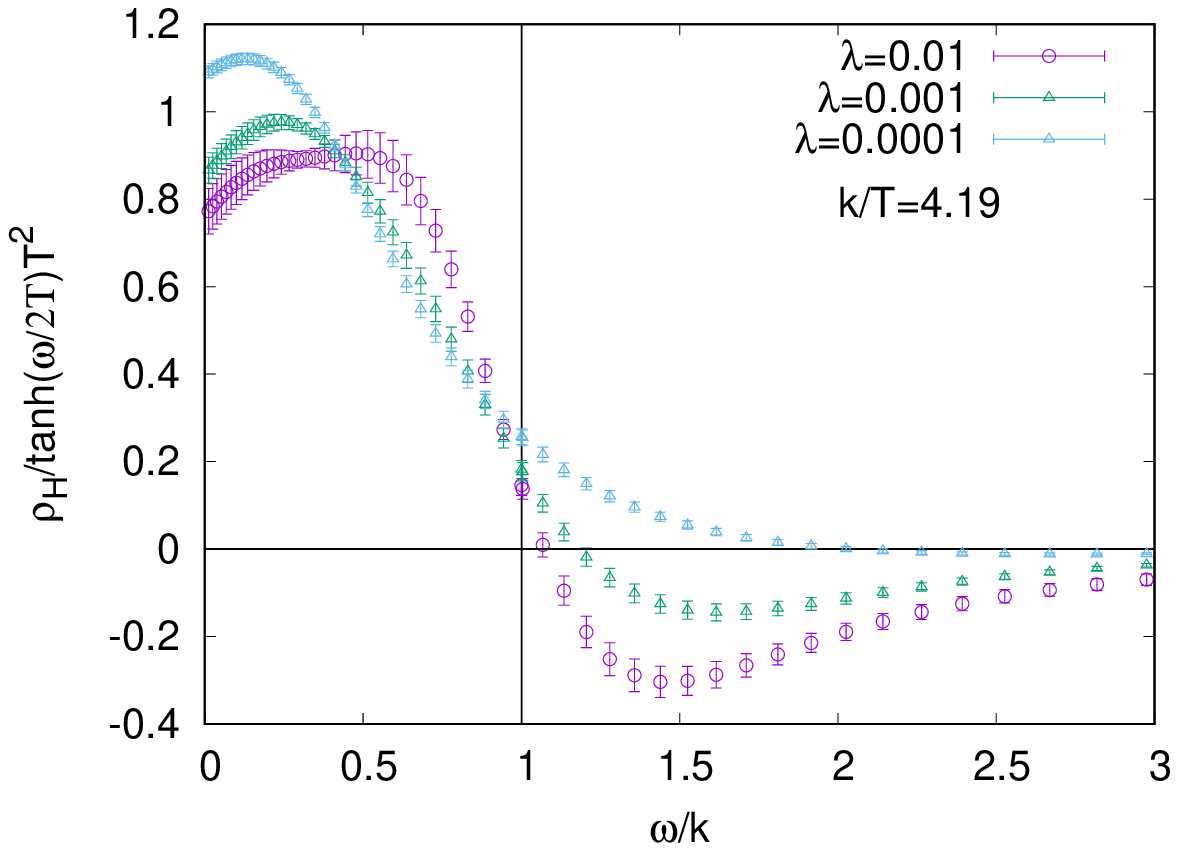}
\includegraphics[width=0.49\textwidth]{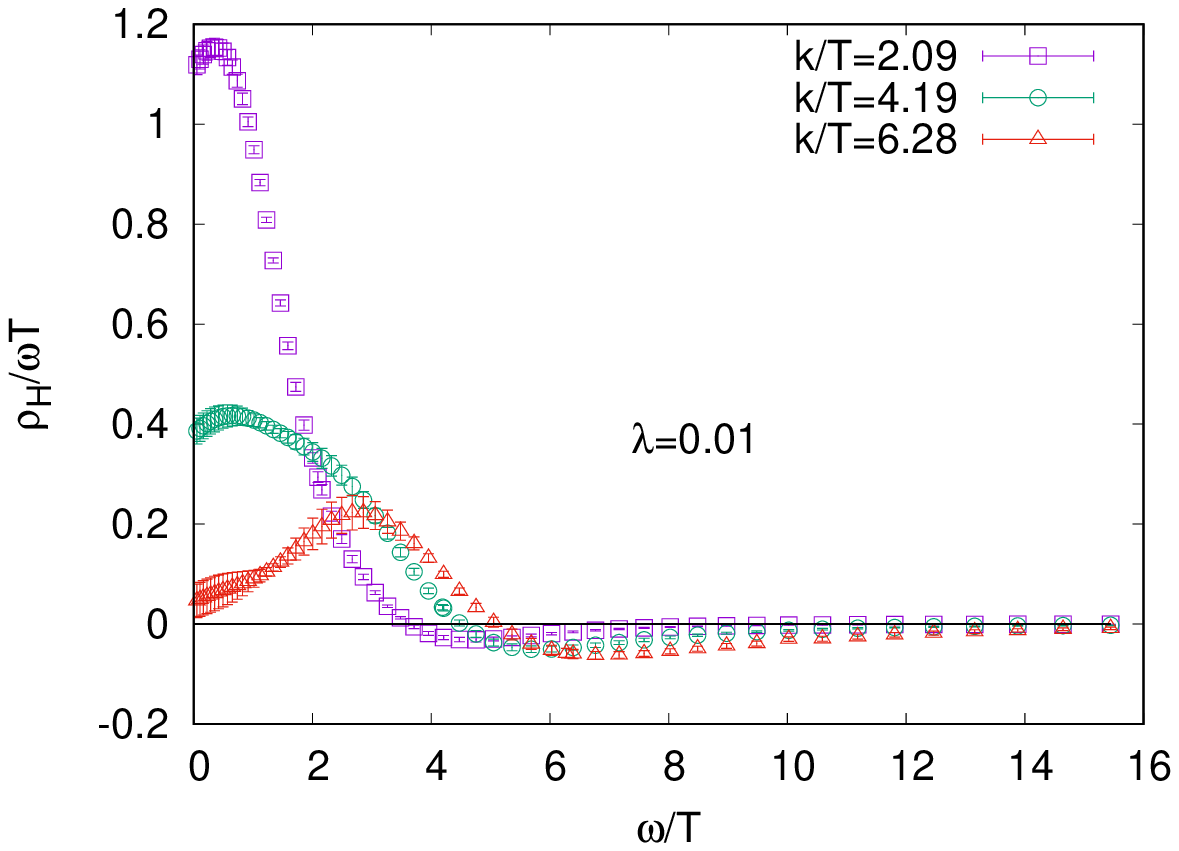}
\caption{Backus-Gilbert estimated spectral function, (Left) Variation of spectral function with $\lambda$ for $k/T=4.19$. (Right) Spectral function for $\lambda=0.01$ for various momenta.
}
\label{bgm_spectral_function}
\end{figure}
\subsection{Backus-Gilbert estimate of $\rho_{H}(\omega,\vec k)$}
The third approach that we have used is the Backus-Gilbert (BG) method~\cite{BG}. BG method has been widely used for the spectral reconstruction from lattice correlator ~\cite{BFMR, BFJM}. In this method, one rewrites the integral transform (\ref{unstabel}) in the following form,
\begin{equation}
G_{H}(\tau, \vec k) \; = \;
\int_{0}^{\infty} d\omega \,
\frac{\rho(\omega,\vec k)}{ f(\omega,\vec k)} \, 
K(\omega,\tau) \; ; \quad
K(\omega,\tau)\equiv\frac{f(\omega,\vec k) \cosh[\omega(\beta/2-\tau)]}{\sinh(\omega \beta/2)\pi}\, . 
\label{BG}
\end{equation}
The purpose of the function $f(\omega,\vec k)$ is to put physics information in the reconstruction process and remove the singularity originating from the denominator of the kernel.

The BG strategy is to write the spectral function 
as a linear combination of the temporal lattice correlator, 
($i$ enumerates the measured points) 
\begin{equation}
    \frac{\rho_{H}^{\rm BG}(\omega,\vec k)}{f(\omega,\vec k)}
    \; = \; 
    \sum_{i} q_{i}(\omega,\vec k) \,G_{H}(\tau_{i},\vec k)\\
    \; = \;
    \int_{0}^{\infty} d \omega^{\prime} 
    \delta(\omega^{\prime},\omega) 
    \frac{\rho_{H}(\omega^{\prime},\vec k)}{f(\omega^{\prime},\vec k)} \, ,
\end{equation}
where $\delta(\omega,\omega^{\prime})=\sum_{i} q_{i}(\omega,\vec k) K(\omega^\prime,\tau_i)$
is the so-called resolution function.
The BG estimated spectral function would be the exact spectral function when $\delta(\omega,\omega^{\prime})=\delta(\omega-\omega^{\prime})$. Therefore the goal is to minimize the width of the resolution function as small as possible by varying $q_{i}$, which can be done by minimizing the following function,
\begin{equation}
A(\omega) \; = \;
\int d\bar \omega(\omega-\bar \omega)^2 \delta(\omega,\bar \omega)^2
    \; = \;
    {q^{t}(\omega).W(\omega).q(\omega)} \, ,
\end{equation}
where  $W_{ij}(\omega)\equiv \int_0^{\infty} d\bar \omega \, (\omega-\bar\omega)^2 K(\bar \omega,\tau_i) K(\bar \omega, \tau_j)$. 
Minimization of this function requires the inversion of the matrix $W$. However, the $W$  matrix is badly conditioned, which causes the minimization to be unstable. 
The minimization of the following regularized function can overcome this,
\begin{equation}
F(\omega)=\lambda A(\omega)+(1-\lambda) B(\omega)
\end{equation}
where $B(\omega)= {\rm Var}[\rho^{\rm BG}(\omega)]={q^{t}(\omega).S.q(\omega)}$ with $S$ being the covariance matrix. Various other regularization schemes have also been used in literature, for example, one such regularization is the Tikhonov scheme where $S_{ij} = \delta_{ij}$ \cite{ABK}.   
The parameter $\lambda$ is related to the width of the resolution function and as well as the error on the spectral function. The `better' choices of $f(\omega,\vec k)$ should also help reduce the resolution function's width.

For the calculation of the BG estimated spectral function for our lattice correlator, we used the following function,
\begin{equation}
f(\omega) \; = \; 
\frac{\tanh(\omega/T)}{1+(\omega/\omega_0)^2+(\omega/\omega_0)^4}
\end{equation}
where, $\omega_{0}=\sqrt{k^2+\pi^2 T^2}$.
This functional form is motivated by the observation that at small $\omega$, the spectral function should be proportional to $\omega$, whereas, at large $\omega$, the spectral function should go like $1/\omega^4$.

\begin{figure}
\centering
\includegraphics[width=0.49\textwidth]{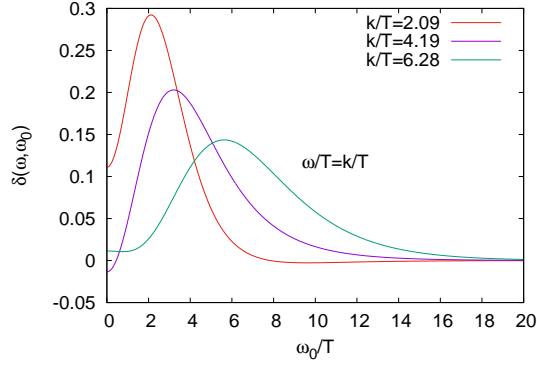}
\caption{Resolution function obtained from the BG method on the light cone for various momenta for $\lambda=0.01$.
}
\label{res_function}
\end{figure}
The BG estimated spectral function is plotted for various values of $\lambda$ in the left panel of Fig.~\ref{bgm_spectral_function}. The variation with respect to $\lambda$ is taken to the systematic error. The spectral function for various momenta for $\lambda=0.01$ is shown in the right panel of Fig.~\ref{bgm_spectral_function}. We observe the same qualitative trend of the spectral function as from the model fit of the data. 
The resolution function at the photon point is plotted in Fig.~\ref{res_function} for all available momenta. We note that the resolution function is peaked around $\omega_0 \simeq |\vec k|\,$.
\begin{figure}
\centering
\includegraphics[width=0.49\textwidth]{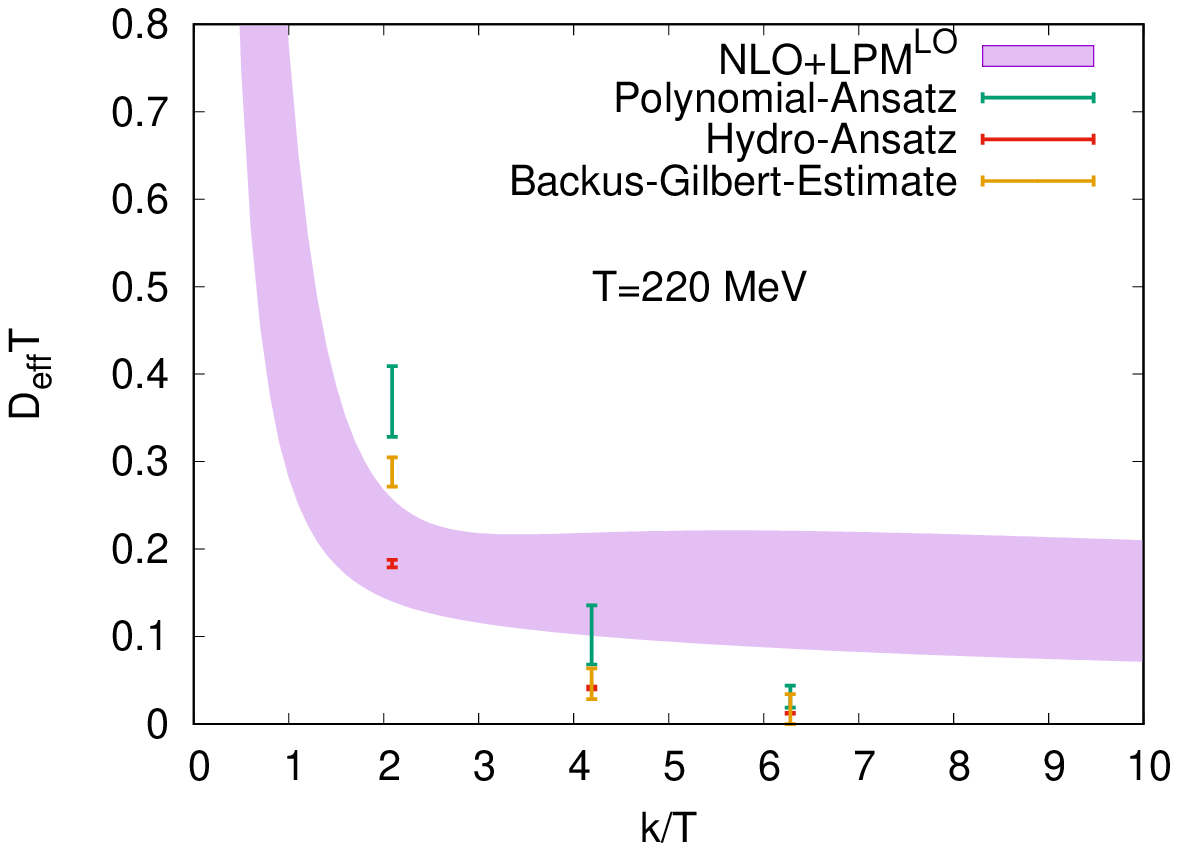}
\includegraphics[width=0.49\textwidth]{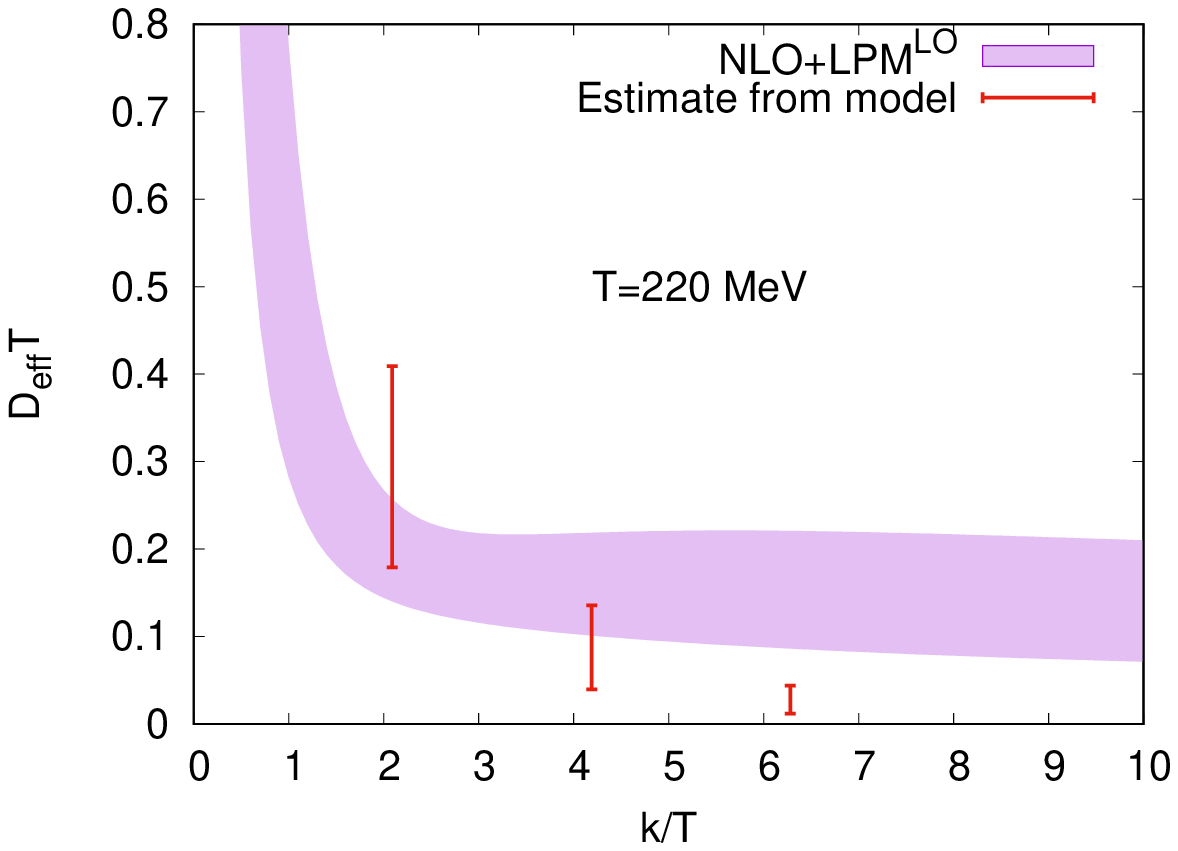}
\caption{(Left) Effective diffusion coefficient obtained from various methods.
(Right) Effective diffusion coefficients obtained from various models have been combined to get a systematic error band. 
In both panels, the perturbative band is give by varying the scale choice in the running coupling -- see Ref.~\cite{JL} for details. 
}
\label{DEFF}
\end{figure}
\section{Effective diffusion coefficient}
\label{FR}
The photon production rate can be rewritten in terms of the so-called effective diffusion coefficient,
\begin{equation}
D_{\rm eff}(k) \; \equiv \; 
\frac{\rho_H(k,\vec k)}{2\chi_{q}k} \, .
\end{equation}
The value of $D_{\rm eff}(k)$ estimated from two models, and the BG method is shown in the left panel of Fig.~\ref{DEFF}. 
We see that at small momentum, the model prediction has considerable uncertainty, whereas, at large momentum, different methods seem to agree mutually. 
We also observe that at large momentum, $D_{\rm eff}$ is smaller than the perturbative estimate. In the right panel, we combined the model results to get a systematic uncertainty on $D_{\rm eff}$.

\section{Conclusion}
We estimated the photon production rate from the $T-L$ correlator from $N_f=2+1$ flavor QCD for $m_{\pi}=313$ MeV at $T=220$ MeV. 
The $T-L$ correlator was calculated using Clover-improved Wilson fermions on HISQ configurations. Various models have been used for spectral reconstruction from the lattice correlator. For the first model, we have used a 
piecewise 
polynomial 
which smoothly connects 
a quintic function of $\omega$ in the IR with 
the asymptotic $1/\omega^4$ form in the UV. 
We apply this model first to  a mock correlator from NLO+LPM 
resummed 
perturbation theory, where we know the exact spectral function. 
Within systematic uncertainty, we could reproduce this perturbative spectral function.
Then we fit this model to the lattice correlator,
with the result 
for various momenta 
shown in the left panel of Fig.~\ref{spectral-function-variation-M}. 
A model based on relaxation time hydrodynamics was similarly considered,
with the resulting spectral function being shown in the right panel of Fig.~\ref{spectral-function-variation-M}. 
We have also used the Backus-Gilbert method to estimate the spectral function. In all these methods, we found qualitative agreement; namely, the spectral function is IR-dominated. 
The 
effective diffusion coefficient 
from all these methods is shown in Fig.~\ref{DEFF}. 
We use the spread in values obtained via the different methods as a proxy for the systematic uncertainty from reconstructing the spectral function. 

\section{Acknowledgement}
We thank Luis Altenkort and Hai-Tao Shu for the production of gauge configurations and implementation of meson measurements in the QUDA code. 
This work is supported by the Deutsche Forschungsgemeinschaft (DFG, German 
Research Foundation)-Project number 315477589-TRR 211. For the computations, we used 
high-performance facilities from the Bielefeld and Juelich machines.
G.~J. was supported by the U.S. Department of Energy (DOE) under grant No.~DE-FG02-00ER41132. 
A.~F. acknowledges support by the Ministry of Science and Technology Taiwan (MOST) under grant 111-2112-M-A49-018-MY2.

\end{document}